\documentclass[reprint, twocolumn, amsmath, amssymb, aps, showpacs, prl]{revtex4-1}

\usepackage{graphicx}
\usepackage{dcolumn}
\usepackage{bm}

\begin{document}

\title{Comment on article ``Line of Dirac Nodes in Hyperhoneycomb Lattices''}
\author{Marcos Ver\'{\i}ssimo-Alves}
\email{Corresponding author: marcos\_verissimo@id.uff.br}
\author{Rodrigo G. Amorim}%
\author{A. S. Martins}
\affiliation{%
Departamento de F\'{\i}sica, ICEx, Universidade Federal Fluminense, Volta Redonda, RJ, Brazil
}
\date{\today}
\maketitle

Recently, carbon hyperhoneycomb ($\mathcal{H}$-n) lattices have 
been proposed to be, through a one-$p$ orbital tight-binding (TB) 
model with nearest-neighbor (n.n.) hopping, a family of strong 
3D topological insulators \cite{ref:mullen}. In this Comment we 
show, through Density Functional Theory (DFT) and Extended H\"uckel 
Theory (EHT) calculations \cite{ref:parameters} for the $\mathcal{H}$-0 
lattice, that the TB Hamiltonian in Ref. \cite{ref:mullen} 
is inadequate, invalidating the presented conclusions.

\begin{figure}[h]
\centering
\includegraphics[width=0.4\textwidth]{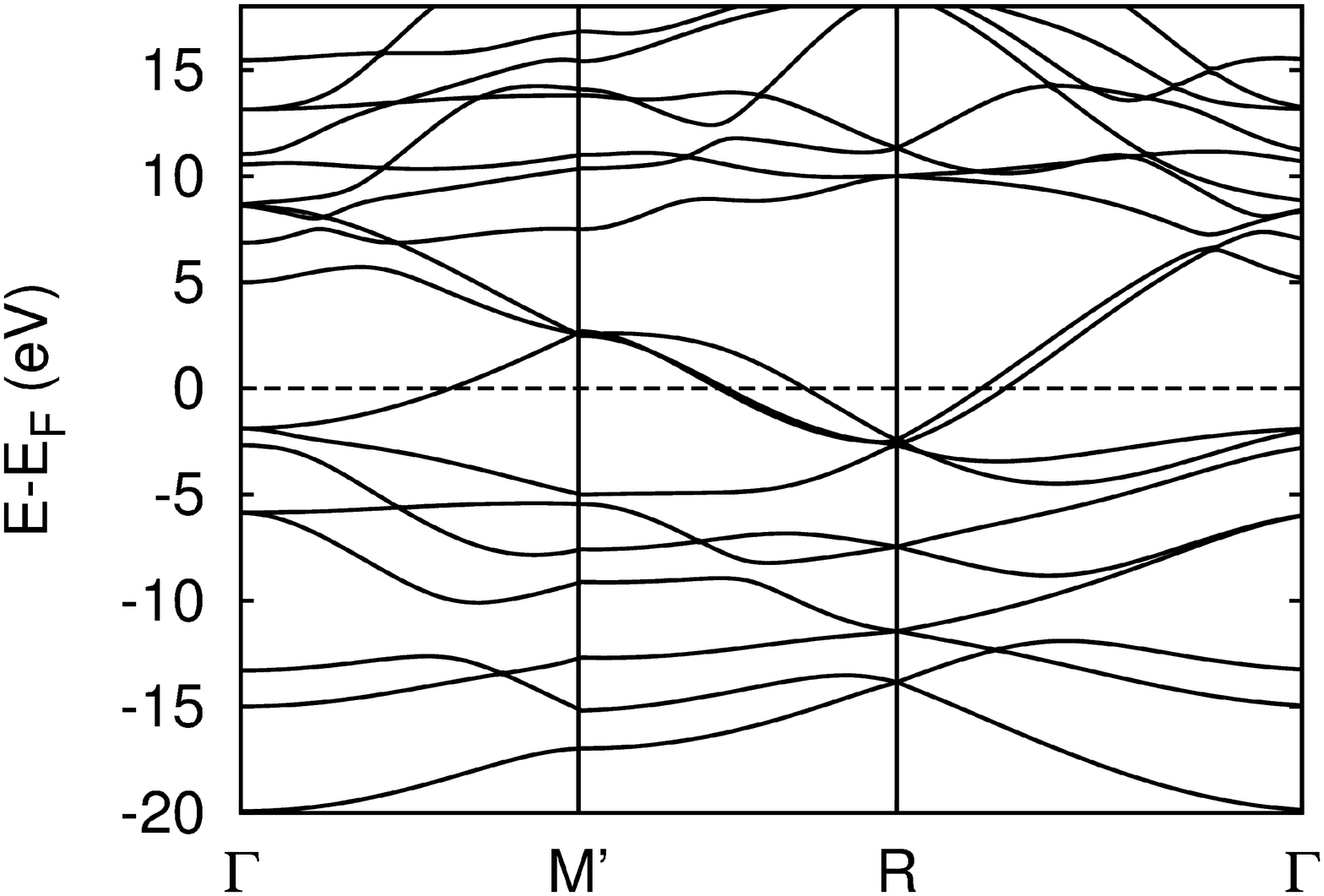}
\caption{DFT band structure for the $\mathcal{H}$-0 lattice.} 
\label{fig:h0_dft_bands}
\end{figure}

Fig. \ref{fig:h0_dft_bands} shows the DFT band structure for the
$\mathcal{H}$-0 lattice. The strong discrepancy between DFT and TB
band structures led us to carefully re-examine the TB Hamiltonian of 
Ref. \cite{ref:mullen}. Using Fig. 1 (a) of
Ref. \cite{ref:mullen} to denote the four atoms in the unit cell, we
rewrite the TB Hamiltonian with different $t_{\alpha \beta}$ between 
atoms $\alpha$ and $\beta$, obtaining:

\begin{equation}
\mathcal{H}_{\alpha \beta} = \left(
\begin{array}{c c c c}
0                & t_{12}\Theta_x & 0                & t_{14}\Theta_z \\
t_{12}\Theta_x^* & 0              & t_{23}\Theta_z^* & 0 \\
0                & t_{23}\Theta_z & 0                & t_{34}\Theta_y \\
t_{14}\Theta_z^* & 0              & t_{34}\Theta_y^* & 0
\end{array}
\right ),
\label{eq:ham_TB_gen}
\end{equation}
where $\Theta_z=\exp{(ik_za/2)}$ and $\Theta_i=2\Theta_z \cos{\sqrt{3}k_ia/2}$, 
$i=x,y$. For $t_{\alpha \beta}=t \mbox{ }\forall\mbox{ } \alpha, \beta$, 
Eq. \ref{eq:ham_TB_gen} reduces to the Hamiltonian of 
Ref. \cite{ref:mullen}.

\begin{figure}[htb]
\centering
\includegraphics[width=0.4\textwidth]{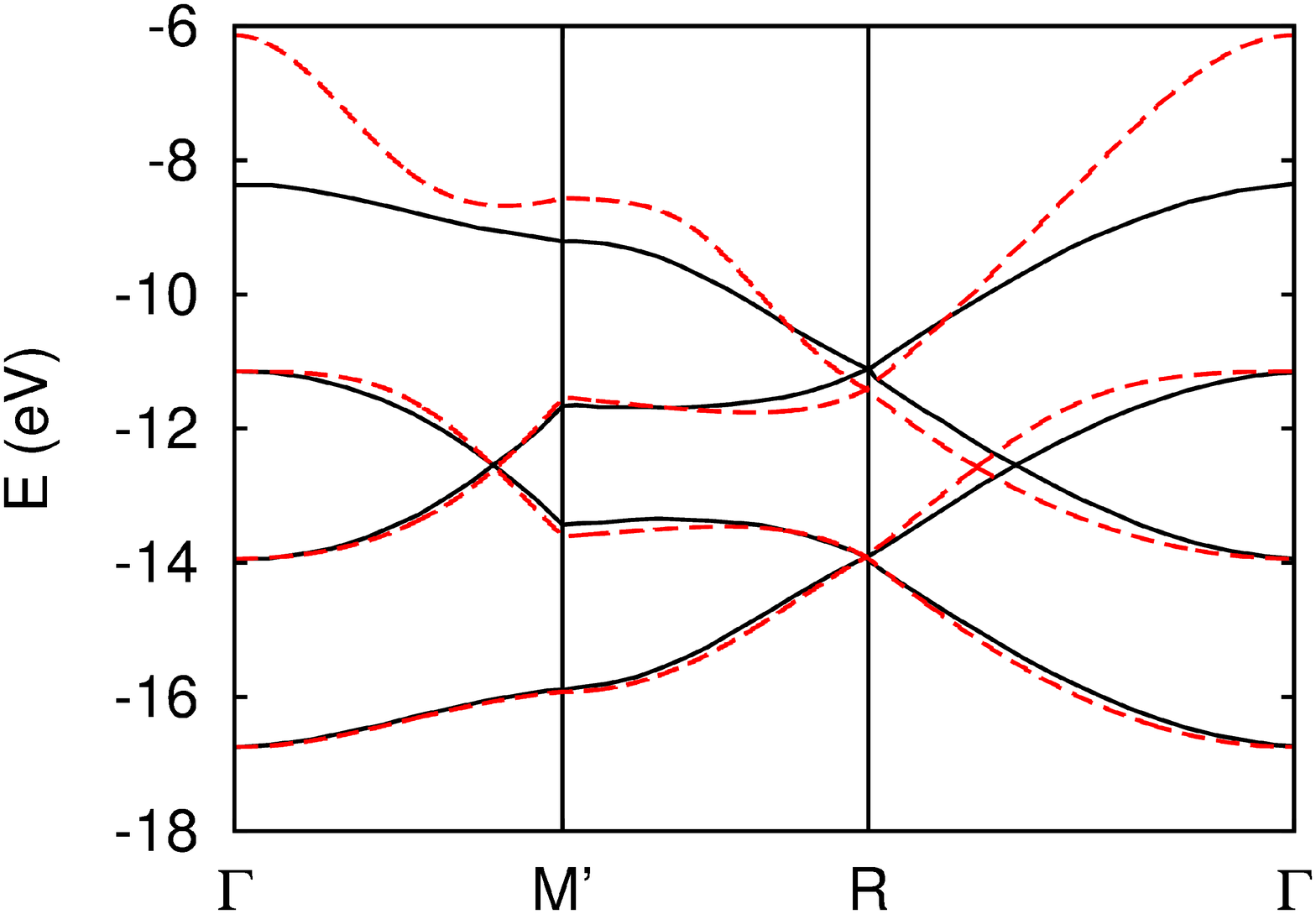}
\caption{(Color online) Comparison between lowest-lying H\"uckel bands (red, dashed lines) and TB bands (solid black lines) of Ref. \cite{ref:mullen}.}
\label{fig:superp_huckel_TB}
\end{figure}

In the $\mathcal{H}$-0 lattice, atom pairs 1-2 and 3-4 lie on
perpendicular planes. For these pairs, $p$ orbitals will be
identical and parallel yielding $t_{12}=t_{34}=t_{\pi}$. $p$ 
orbitals on pairs 1-4 and 2-3, however, will be orthogonal, 
hence $t_{23}=t_{14}=0$. Now the Hamiltonian (\ref{eq:ham_TB_gen}) 
reads

\begin{equation}
\mathcal{H}_{\alpha \beta} = t_{\pi}\left(
\begin{array}{c c c c}
0          & \Theta_x & 0          & 0 \\
\Theta_x^* & 0        & 0          & 0 \\
0          & 0        & 0          & \Theta_y \\
0          & 0        & \Theta_y^* & 0
\end{array}
\right ).
\label{eq:correct_ham_TB}
\end{equation}
Since $t_{14}$ and $t_{23}$ can only be non-zero if the orbitals 
have spherical symmetry, the Hamiltonian in Ref. \cite{ref:mullen} 
is effectively built from $s$, and not $p$, orbitals. This is 
confirmed by the comparison of the lowest-lying 1$^{st}$ n.n. 
H\"uckel bands, which have a strong $s$-character, and TB bands 
of Ref. \cite{ref:mullen}, shown in Figure \ref{fig:superp_huckel_TB}. 

From Eq. (\ref{eq:correct_ham_TB}) the zero-energy eigenvalues 
are now planes $k_{x(y)}=\pm \frac{2 \pi}{a\sqrt{3}} \mbox{ } \forall \mbox{ } k_{y(x)},k_z$, 
ruling out the existence of lines of Dirac loops, and 
removing the $\mathcal{H}$-n lattices from the class of 
strong 3D topological insulators, at least within the TB 
approximation used in Ref. \cite{ref:mullen}. 

\begin{acknowledgments} 
The authors acknowledge Profs. B. Koiller, A. L. S. Oliveira, 
R. B. Capaz and A. G. M. Schmidt for enlightening discussions 
and critical reading of this manuscript. M. V. Alves and A. S. 
Martins acknowledge funding agency FAPERJ for financial support 
through grants E-26/111.397/2014 and E-26/112.554/2012. 
\end{acknowledgments}


\begin{thebibliography}{100}
\bibitem{ref:parameters} We carry out DFT calculations \cite{ref:dft} 
with the SIESTA \cite{ref:siesta} code, using Troullier-Martins 
pseudopotentials \cite{ref:tm2}, the PBE exchange-correlation 
functional \cite{ref:pbe} and a 900 Ry mesh cutoff. EHT \cite{ref:huckel} 
calculations employed a double-$\zeta$ $spd$ basis set 
and the parametrization for carbon systems from Ref. \cite{ref:cerda}. 
The Brillouin Zone is sampled with a $18 \times 18 \times 12$ 
Monkhorst-Pack grid \cite{ref:mpack} in both DFT and EHT calculations.
\end{thebibliography}
\end{document}